\documentclass[reprint,aps,amsmath,amssymb,amsfonts,dvipdfmx]{revtex4-1}

\usepackage[dvipdfmx]{graphicx,color}

\usepackage{braket,bm,comment}
\usepackage{dcolumn}
\begin{document}

\title{Skyrmion crystal formation and temperature - magnetic field phase diagram of the frustrated tirangular-lattice Heisenberg magnet with easy-axis masugnetic anisotropy}


\author{Hikaru Kawamura}
\email[]{kawamura@ess.sci.osaka-u.ac.jp}
\affiliation{
	Molecular Photoscience Research Center, Kobe University, Kobe, Hyogo 657-8501, Japan
}

\date{\today}

\begin{abstract}
 The nature of the skyrmion-crystal (SkX) formation and various multiple-$q$ phases encompassing the SkX phase are investigated by extensive Monte Carlo simulations on the frustrated $J_1$-$J_3$ triangular-lattice Heisenberg model with the weak easy-axis magnetic anisotropy. Phase diagram in the temperature $T$ vs. magnetic-field $H$ plane are constructed, leading to a rich variety of multiple-$q$ phases. The anisotropy stabilizes the SkX state down to $T=0$ at intermediate fields, while in the lower-field range the SkX state becomes only metastable, and new multiple-$q$ states with a broken $C_3$ symmetry are instead stabilized. Implications to experiments are discussed.
\end{abstract}


\maketitle

Much attention has recently been paid to various types of topologically protected nanoscale spin textures in magnets, e.g., vortex, skyrmion and hedgehog, from both fundamental interest in topology-related physics and possible applications to spintronics. Skymion, a swirling noncoplanar spin texture characterized by an integer topological charge whose constituent spin directions wrap a sphere in spin space, has got special attention. In magnetically ordered state, skyrmion is often stabilized as a periodic array called the skyrmion crystal (SkX). At an earlier stage, the SkX state was discussed  for non-centrosymmetric magnets as induced by the antisymmetric Dzyaloshinskii-Moriya (DM) interaction \cite{Muhlbauer,Neubauer,Munzer,Yu2010,Yu2011}. In 2012, it was theoretically proposed  that the ``symmetric'' SkX is also possible in certain class of {\it frustrated\/} centrosymmetric magnets without the DM interaction, where the size of constituent skyrmions can be varied continuously from very small to infinitely large (corresponding to the continuum limit) by tuning the extent of frustration \cite{OkuboChungKawamura}. An interesting characteristic of frustration-induced symmetric skyrmion is that, due to the underlying chiral degeneracy, both skyrmion and antiskyrmion of mutually opposite signs of topological charge, or the scalar chirality, are equally possible, leading to the unique and rich electromagnetic responses \cite{OkuboChungKawamura}.

 In Ref.\cite{OkuboChungKawamura}, the SkX was identified 
in a simplified model, i.e., the frustrated $J_1$-$J_3$ ($J_1$-$J_2$) isotropic Heisenberg model on the triangular lattice as a triple-$q$ state stabilized by magnetic fields and thermal fluctuations. Subsequent experiment successfully observed the SkX for centrosymmetric triangular-lattice metallic magnet, e.g., Gd$_2$PdSi$_3$, accompanied by the pronounced topological Hall effect \cite{Kurumaji}. Recent Monte Carlo (MC) simulation indicated that the SkX could also be stabilized in the standard RKKY system with only the bilinear interaction modelling weak-coupling metals, where the oscillating nature of the RKKY interaction bears frustration \cite{Mitsumoto2021, Mitsumoto2022}.

  Of course, real material possesses various perturbative interactions not taken into account in a simplified model \cite{OkuboChungKawamura}, 
e.g., the three-dimensionality (interplane coupling), the magnetic anisotropy, quantum fluctuations, etc. In particular, experiment has indicated that the SkX can be stabilized even at zero temperature ($T=0$), where the effect of certain perturbative interactions, e.g., the magnetic anisotropy, was argued to play a role \cite{Kurumaji, LeonovMostovoy}. Possible mechanism leading to the $T=0$ SkX state was theoretically discussed in the literature, including the biquadratic interaction arising from the higher-order perturbation beyond the second-order (strong-coupling effect in itinerant metals) \cite{OzawaHayamiMotome2017, HayamiOzawaMotome2017, WangBatista}, quantum spin fluctuations \cite{Rosch}, etc. 
Among them, the magnetic anisotropy prevails in real magnets, both classical and quantum, and generally exists even at the spin-bilinear order.

 On the basis of the ground-state phase diagram of the frustrated $J_1$-$J_2$ triangular Heisenberg model obtained by the simulated annealing, it was theoretically suggested that the easy-axis magnetic anisotropy stabilized the SkX state even at $T = 0$ \cite{LeonovMostovoy}. 
While the effect of magnetic anisotropy on the SkX formation was examined further by various authors \cite{LinHayami2016,HayamiLinBatista, HayamiMotome2019,Wang2020, Wang2021,HayamiXY,Hayami2022,Rosch}, 
most of them concentrated on the $T=0$ properties, with few studies on 
the temperature ($T$) vs. magnetic-field ($H$) phase diagram (see \cite{LinHayami2016,HayamiLinBatista,HayamiXY}, however). 
Even concerning with the $T=0$ properties, 
the proposed magnetic-anisotropy stabilization of the SkX state might deserve further careful examination, since the numerical method employed, e.g., the simulated annealing, might capture the metastable SkX state, while such a metastable, not truly stable SkX state was indeed reported under certain annealing conditions even experimentally \cite{Oike,Karube}. 

 Under these circumstances, we study by extensive MC simulations the SkX formation and the $T$-$H$ phase diagram of the frustrated $J_1$-$J_3$ Heisenberg model on the triangular lattice with the easy-axis magnetic anisotropy, an anisotropic extention of the isotropic model of Ref.\cite{OkuboChungKawamura}. We wish to clarify how the $T$-$H$ phase diagram of the isotropic model changes by the magnetic anisotropy, paying special attention to the questions of whether the SkX state is truly stabilized at $T=0$, whether some new phases appear induced by the anisotropy, and if any, the nature of these phases. We then find that the SkX state is stabilized in intermediate fields at $T=0$, while its stability range is considerably reduced compared with that obtained by the simulated annealing, and in the region where the SkX phase turns out to be only metastable two new anisotropy-induced multiple-$q$ phases with a broken $C_3$ symmetry emerge as stable phases.

  We consider the $J_1$-$J_3$ classical Heisenberg model on the two-dimensional triangular lattice with the easy-axis uniaxial anisotropy. The Hamiltonian is given by
\begin{eqnarray}
&\mathcal{H}& = -J_1\sum_{\Braket{i,j}_1}(S_{ix}S_{jx} + 	S_{iy}S_{jy} +\gamma S_{iz}S_{jz}) \nonumber \\
 - &J_3& \sum_{\Braket{i,j}_3}(S_{ix}S_{jx} + S_{iy}S_{jy} +\gamma S_{iz}S_{jz}) - H \sum_i S_{iz} , 
\label{eq:hamiltonian}
\end{eqnarray}
\noindent
where $J_1>0$ is the ferromagnetic nearest-neighbor coupling,  $J_3<0$ the antiferromagnetic third-neighbor coupling, $ {\bm S}_i=(S_{ix}, S_{iy}, S_{iz})$ a three-component unit vector at site $i$, magnetic field is applied along the easy axis with $H$ the magnetic-field intensity, and $\gamma$ the uniaxial exchange anisotropy parameter. We assume a rather weak easy-axis anisotropy and set $\gamma=1.1$, i.e., 10\% anisotropy. 
Following Ref.\cite{OkuboChungKawamura}, we set $J_1/J_3=-1/3$, and $J_1$, $T$ and $H$ are given in units of $|J_3|$,  hereafter.

 MC simulation based on the standard heat-bath method combined with the over-relaxation method is performed. In addition, fully equilibrated temperature-exchange simulations are also made in the higher-$T$ range. The lattice is a $L\times L$ triangular lattice with $L=144,180, 216, 288$ with periodic boundary conditions. Unit MC step consists of one heat-bath and $L$ over-relaxation sweeps. Typically, each run contains $2\times 10^5$ MC steps per spin (MCS) at each temperature, the first half discarded for thermalization.

 To reach a given $(T^*, H^*)$ state, together with the field-cooling (FC) run, i.e., the gradual cooling simulated-annealing run at fixed $H^*$, various other computation protocols are tried by combining $H$- and $T$-sweeps in search for the stable state. Since, at sufficiently low $T=T_0$, a truly stable state should have the lowest energy among several metastable states generated by different protocols, it can be determined by comparing their energies. One standard protocol might be the zero-field cooling (ZFC) run to ($T_0, H^*$), i.e., gradual cooling in zero field (or weak fields of $H\lesssim 1.5$) to a low $T=T_0$ (we set here $T_0=0.1$, 0.05) followed by the gradual increase of $H$ to $H^*$ at $T=T_0$. Such ZFC runs are repeated 10$\sim $20 times in search for the lower-energy state by changing the random numbers and the way of $H$ application. If the ZFC protocol yields a stable state with the lowest energy at ($T_0, H^*$), gradual warming run from that state is also performed to higher $T=T^*$ at fixed $H^*$ (sometimes further cooling run also made). Consistency is then checked by confirming the obtained state to be compatible with that obtained by the $T$-exchange simulations at moderately high $T$.

\begin{figure}[t]
	\centering
	\begin{tabular}{c}
		\begin{minipage}{\hsize}
			\includegraphics[width=\hsize]{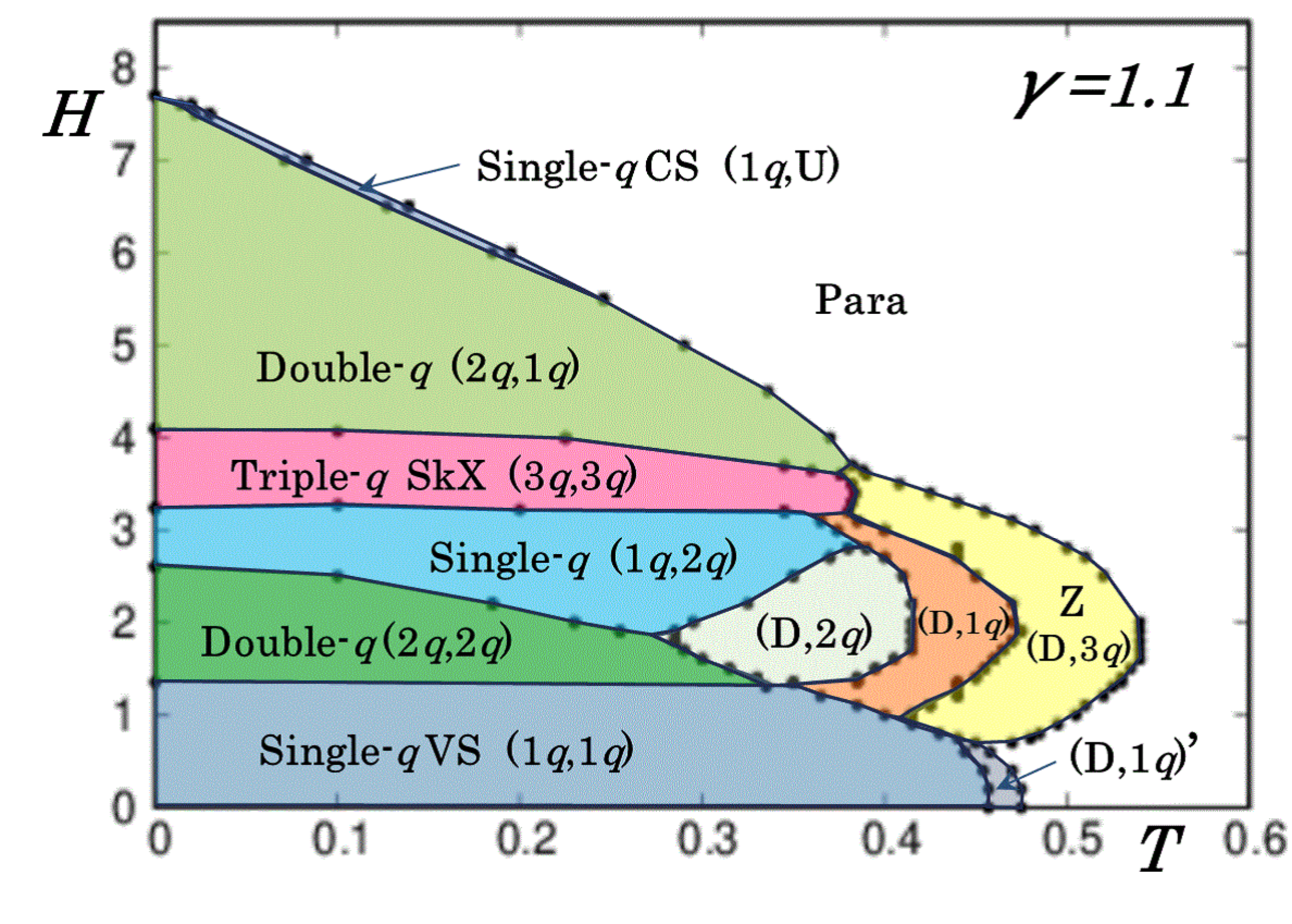}	
		\end{minipage}
	\end{tabular}
	\caption{
 The phase diagram of the model with $J_1/J_3=-1/3$ and $\gamma=1.1$ in the temperature vs. magnetic-field plane. The notations ($mq,nq$), D and U are explained in the text.
	}
	\label{fig1}
\end{figure}

 The $T$-$H$ phase diagram obtained in this way is shown in Fig.1. It contains ten distinct ordered phases. Although it might look rather complicated, all the phases appearing in the isotropic phase diagram \cite{OkuboChungKawamura} also appear here. The triple-$q$ SkX state, described as ($3q,3q$) in Fig.1, is stabilized down to $T=0$ in the intermediate field range, where $m$ and $n$ (=1,2,3) in ($mq, nq$) represent the number of the dominant (quasi-)Bragg peaks except for ${\bf q}=0$ in the transverse ($S_{xy}$) and longitudinal ($S_z$) spin structure factors, $S_\perp({\bm q})$ and $S_\parallel({\bm q})$ \cite{mqnq}. The SkX state is essentially of the same type as the one of the isotropic model, as demonstrated in the real-space spin and scalar-chirality configurations of Figs. 2(c, d). The SkX state is characterized by the nonzero total scalar chirality $\chi_{tot}>0$, leading to the topological Hall effect. The $T$-dependence of the specific heat and $\chi_{tot}$ at $H=3.5$ are shown in Figs. 2(a, b), the definitions of $\chi_{tot}$ and $S({\bf q})$ being given in Appendix A. 
In the higher-$T$ region, the so-called $Z$ phase, the random domain state of the SkX and the anti-SkX \cite{OkuboChungKawamura}, also appears as the collinear triple-$q$ (D,$3q$) state (D means disordered, i.e., the absence of sharp (quasi-)Bragg peak in $S({\bm q})$): See Fig. 11.

\begin{figure}[t]
	\centering
	\begin{tabular}{c}
		\begin{minipage}{\hsize}
			\includegraphics[width=\hsize]{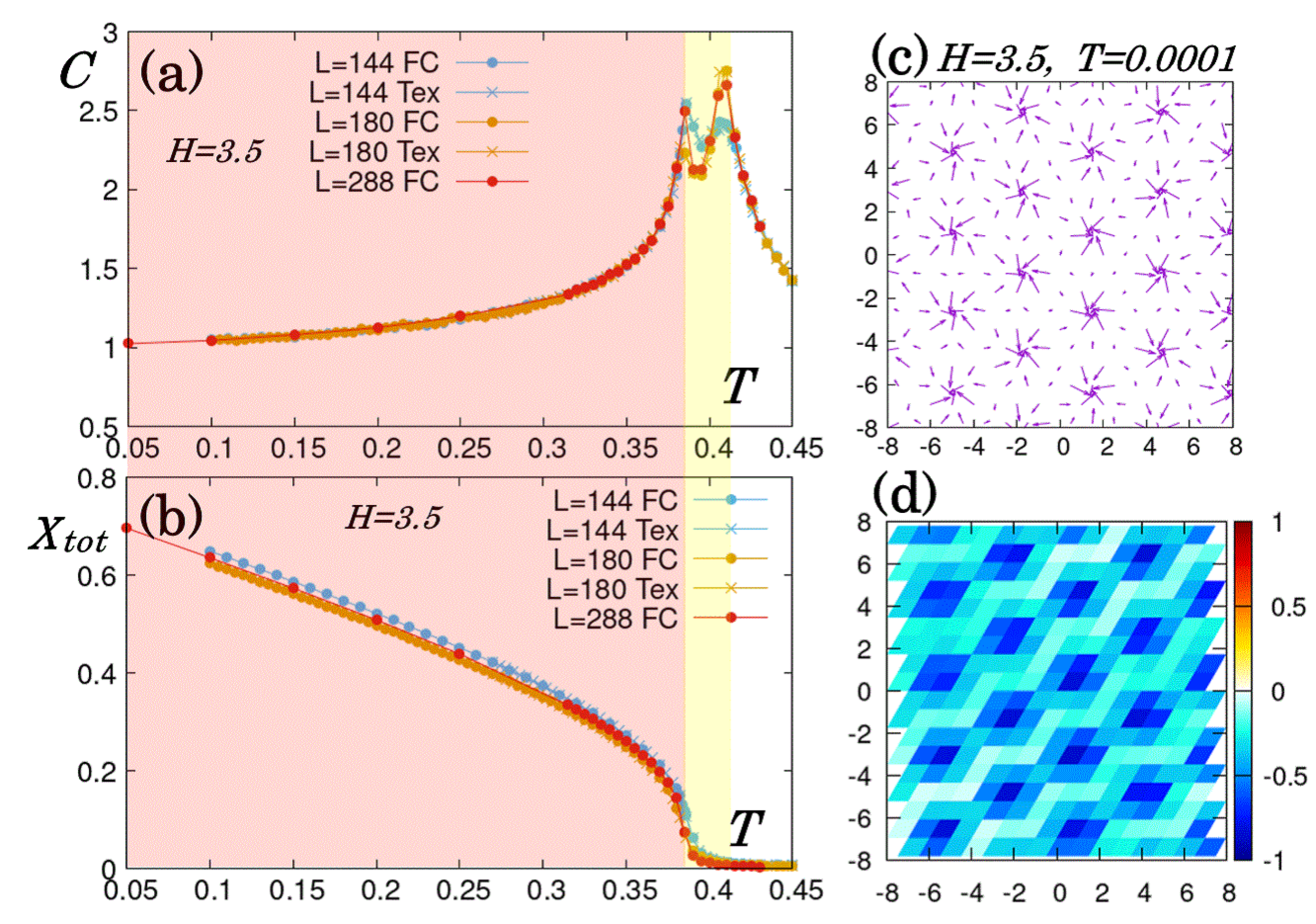}	
		\end{minipage}
	\end{tabular}
	\caption{
 The temperature and size dependence of (a) the specific heat, and of (b) the total scalar chirality, at $H=3.5$. The data of the FC runs and of the temperature-exchange runs ($T_{{\rm ex}}$) are shown. The typical real-space configurations of (c) the transverse components of the spin, and of (d) the scalar chirality, in the SkX state at $H=3.5$ and $T=0.0001$ are shown for a common part of the $L=180$ lattice.
	}
	\label{fig2}
\end{figure}

 The single-$q$ spiral states also appear both in zero (or sufficiently weak) field and in higher fields. 
The first type, the ($1q,1q$) state, is a vertical spiral (VS) induced by the easy-axis anisotropy, exhibiting a $90^\circ$ rotation from the conical spiral (CS) stabilized in the isotropic model \cite{OkuboChungKawamura}. In zero field, this VS state in the $T\rightarrow 0$ limit is vertically coplanar, but it becomes weakly noncoplanar in nonzero fields, the latter corresponding to the ``M state'' of Ref.\cite{LeonovMostovoy}. By contrast, the high-field single-$q$ state, the ($1q$, U) state (U means uniform, i.e., only the ${\bm q}={\bm 0}$ peak in $S_\parallel(\bm{q})$), is essentially the same CS as that of the isotropic model \cite{OkuboChungKawamura}. In the high-field region, there also appears the double-$q$ state, the ($2q,1q$) state, essentially the same as that of the isotropic model \cite{OkuboChungKawamura}. Further details of these states are given in Appendices B and C.

 Between the SkX phase at intermediate $H$ and the VS phase at low enough $H$, there appear two new phases  absent in the isotropic model, i.e., the ($1q,2q$) and the ($2q,2q$) phases, which persist even in the $T\rightarrow 0$ limit. The associated $S_\perp(\bm{q})$ and $S_\parallel(\bm{q})$ are respectively given in Figs. 3(a, b) and 3(e, f). Note that ``$1q$'' (``$2q$'') here means $S({\bm q})$ possesses three (quasi-)Bragg peaks, but the $C_3$ symmetry is broken resulting in one (two) pair of higher-intensity peaks \cite{mqnq}. In both phases, $\chi_{tot}$ vanishes, meaning the absence of the topological Hall effect.

 An important caveat might be in order here: If one makes FC simulated-annealing runs in the field range $2.0\lesssim H\lesssim 3.6$ to low $T$, one ends up with the triple-$q$ SkX state even if one makes a very slow cooling. By contrast, if one makes a ZFC run to ($T_0,H^*$) with $H^*$ in the relevant range, one generally ends up with the $C_3$-symmetry broken state. The energy ($e$) comparison indicates that, for the field $1.35\lesssim H\lesssim 2.6$ the ($2q,2q$) state reached by the ZFC run is stable ($e$ is lower than $e$ of the SkX state reached by the FC run by $\sim$0.54\% and by $\sim$0.27\% at $H=2$ and 2.5, respectively, well beyond the typical error bar of order 0.001\%); for $2.6\lesssim H\lesssim 3.2$ the ($1q,2q$) state is stable ($e$ is lower than $e$ of the SkX state by $\sim$0.12\% at $H=3$); but for $3.2\lesssim H\lesssim 4.1$ the ($3q,3q$) SkX state is stable ($e$ is lower than $e$ of the ($2q,2q$)/($1q,2q$) states by $\sim$0.08\% and by $\sim$0.12\% at $H=3.5$ and 4, respectively). In fact, the energy difference between the ($2q,2q$) and ($1q,2q$) states is rather small of order of the error bar, although we have observed a clear phase transition between these two states with varying $T$: See Figs. 4(b) and 6(b).

 Our observation then indicates that the easy-axis anisotropy stabilizes the triple-$q$ SkX state even at $T=0$ at intermediate $H$ ($3.2\lesssim H\lesssim 4.1$), where its stability range is considerably reduced compared with that obtained by the simulated annealing, and the truly stable state in the lower-$H$ region turns out to be the ($1q,2q$) state for $2.6\lesssim H\lesssim 3.2$, and the ($2q,2q$) states for $1.35\lesssim H\lesssim 2.6$. To the author's knowledge, these two states, the ($1q,2q$) and ($2q,2q$) states, are new unnoticed so far. The observed strong hysteretic effect might give the reason why these states were not reported in the $T=0$ phase diagram constructed by, e.g., the simulated annealing \cite{LeonovMostovoy}.

\begin{figure}[t]
	\centering
	\begin{tabular}{c}
		\begin{minipage}{\hsize}
			\includegraphics[width=\hsize]{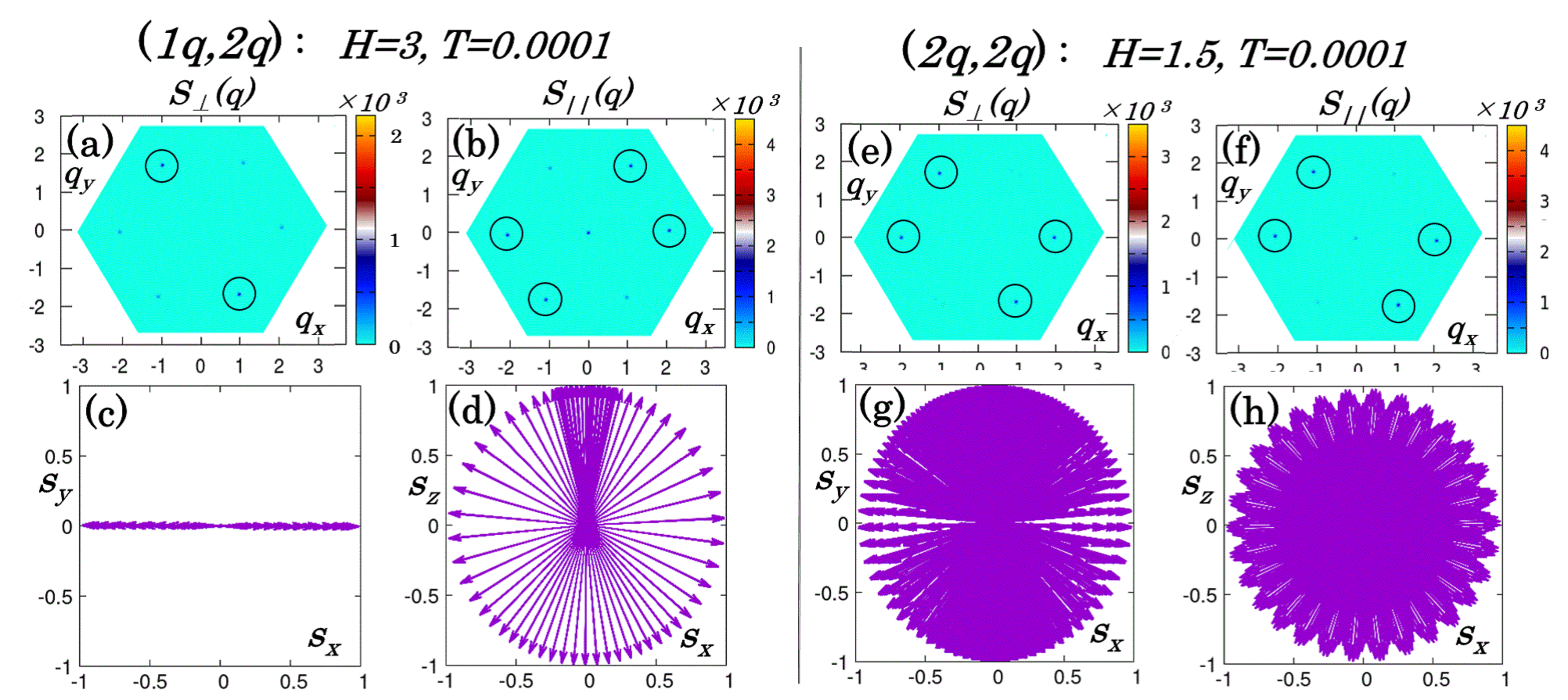}	
		\end{minipage}
	\end{tabular}
	\caption{
 The spin structure factors (a) $S_\perp({\bm q})$, and  (b) $S_\parallel({\bm q})$, and the projected spin (c) ($S_{x},S_{y}$), and (d) ($S_{x},S_{z}$) components, in the ($1q,2q$) phase at $H=3$ and $T=0.0001$. (e)- (h): The corresponding plots in the ($2q,2q$) phase at $H=1.5$ and $T=0.0001$.  The lattice size is $L=180$. 
	}
	\label{fig3}
\end{figure}

 Let us further look into the nature of these new phases.  Although the ($1q,2q$) state is a vertical coplanar state as shown in Figs. 3(c, d) \cite{qandS}, it is not a simple VS ($1q,1q$) state. In fact, as can be seen from the vector-chirality ($\kappa_x,\kappa_y$) projection shown in Fig. 4(c), where ${\bm \kappa}$ is the vector chirality defined on each upward triangle by ${\bm \kappa}= (2/3\sqrt{3})\sum_{<ij>}{\bm S}_i\times {\bm S}_j$ (the summation taken over three clockwise bonds on each triangle) changing sign under the spatial inversion, ${\bm \kappa}$ mostly exhibits parallel alignment in the direction perpendicular to the coplanar spin plane, but some ${\bm \kappa}$ exhibits antiparallel alignment in the opposite direction, indicating that the spins rotate mostly in a certain (say, clockwise) direction, but occasionally rotate in an opposite (say, anti-clockwise) direction. Closer inspection reveals that such a counter-rotation occurs when the spins stay in the vicinity of the $H$-direction to gain the Zeeman energy.

 By contrast, the ($2q,2q$) state is a noncoplanar state as shown in Figs. 3(g, h). As can be seen from the ($\kappa_x,\kappa_z$) projection of Fig. 4(d), ${\bm \kappa}$ is dominated by the horizontal (perpendicular to the field) component, which suggests that the associated noncoplanar spin configuration is basically ``vertical''. If one compares this with the corresponding plot for another double-$q$ state, i.e., the ($2q,1q$) state in high fields, ${\bm \kappa}$ in the latter rather lacks in the horizontal component (see Fig. 4(f)), consistently with the meron-like ``conical'' character of its spin state (see Fig. 4(e)). Hence, the two double-$q$ states, the ($2q,2q$) and ($2q,1q$) states, are different kinds of states, i.e., ``vertical'' vs. ``conical''.

\begin{figure}[t]
	\centering
	\begin{tabular}{c}
		\begin{minipage}{\hsize}
			\includegraphics[width=\hsize]{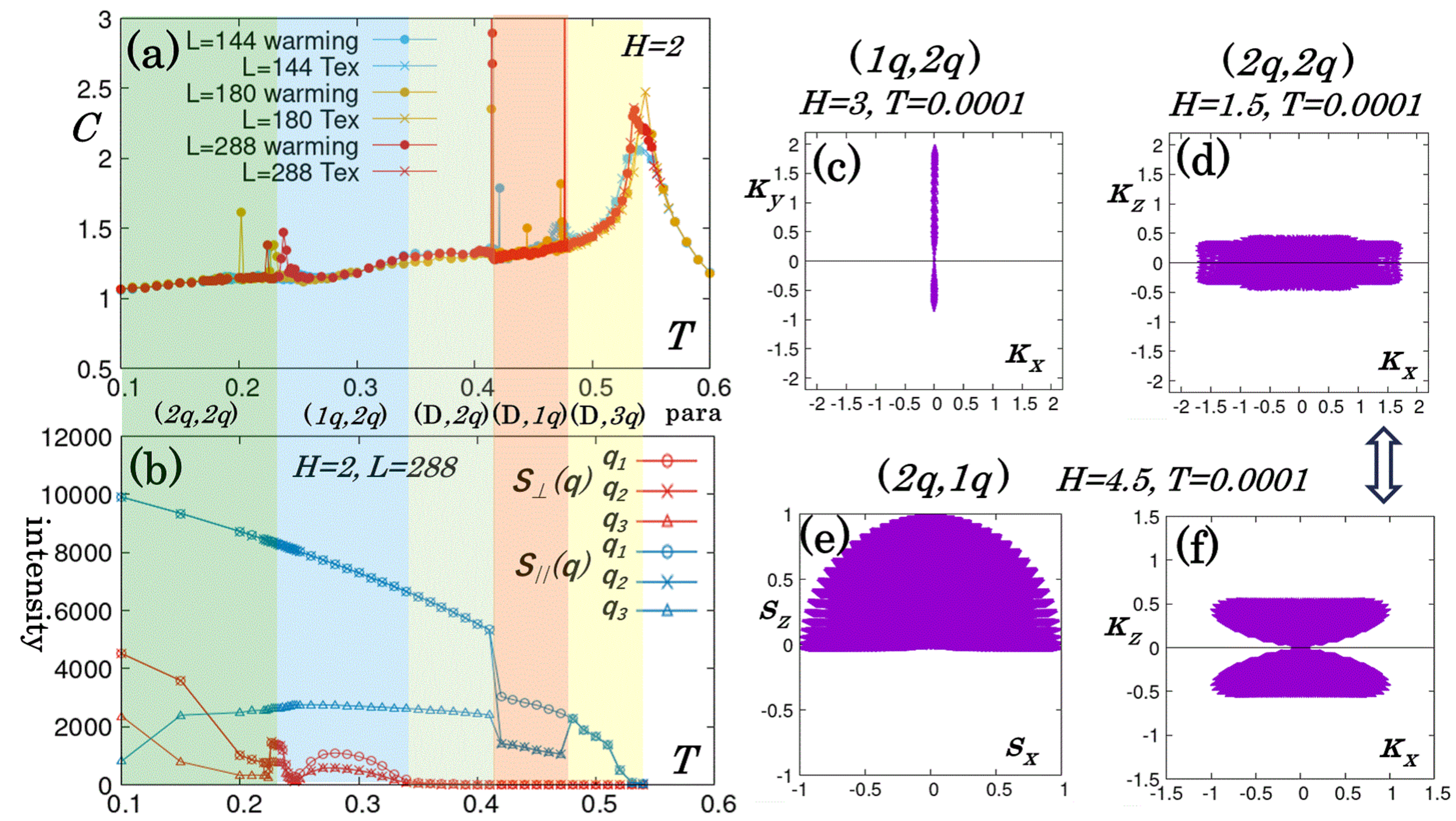}	
		\end{minipage}
	\end{tabular}
	\caption{
The temperature dependence of (a) the specific heat for various sizes, and of (b) the intensities of the three (quasi-) Bragg peaks of 
$S_\perp({\bm q})$ and $S_\parallel({\bm q})$ for $L=288$, at $H=2$. The data are taken by the gradual warming run from the $T=T_0$ state as explained in the text, while in (a) the data of the temperature-exchange run ($T_{{\rm ex}}$) are also added. The projected vector-chirality (c) ($\kappa_{x},\kappa_{y}$) components in the ($1q,2q$) phase at $H=3$ and $T=0.0001$, and (d) ($\kappa_{x},\kappa_{z}$) components in the ($2q,2q$) phase at $H=1.5$ and $T=0.0001$.  The projected (e) spin ($S_{x},S_{z}$), and (f) vector-chirality ($\kappa_{x},\kappa_{z}$) components, in the ($2q,1q$) phase at $H=4.5$ and $T=0.0001$.  
	}
	\label{fig4}
\end{figure}

 In the higher-$T$ range, phases absent in the isotropic model also appear, including the collinear single-$q$ (D,$1q$) 
and the collinear double-$q$ (D,$2q$) phases. In Fig.1, the collinear single-$q$ phase appears in two distinct regions, i.e., at intermediate fields and at zero and weaker fields, each represented by (D,$1q$) and (D,$1q$)', which are not connected in the phase diagram. Indeed, in the (D,$1q$) state, $S_\parallel ({\bm q})$ possesses three pairs of (quasi-)Bragg peaks among which single pair $\pm {\bm q}_1^*$ exceeds the other twos $\pm {\bm q}_2^*$ and $\pm {\bm q}_3^*$ by factor of $2\sim 3$ in their intensities (refer to Figs. 4(b), 6(b) and 10, 
while, in the (D,$1q$)' state,  $S_\parallel ({\bm q})$ possesses only one pair of (quasi-)Bragg peaks. 


 The richness of the phase diagram suggests that even a simple cut of the phase diagram could yield many phases and phase transitions among them. We demonstrate such richness by showing the $T$-dependence of physical quantities at a representative field $H=2$. The data are taken by the gradual warming runs from the $T=T_0$ state prepared by the ZFC run explained above. On increasing $T$ from $T_0$, one encounters the ($2q,2q$), ($1q,2q$), (D,$2q$), (D,$1q$), (D,$3q$) states before finally reaching the paramagnetic state. We show in Fig.4 the $T$-dependences of (a) the specific heat, and of (b) the intensities of the three relevant (quasi-)Bragg peaks ${\bm q}_i^*$ ($i=1,2,3$) of the spin structure factor which are ordered  according to their intensities \cite{comment}. Similar data for $H=2.5$ are also given in Fig. 6. 
As can be seen from the figures, the system indeed exhibits a rich phase structure.

  In view of the complicated appearance of the phase diagram, we now try to give a rough and intuitive picture of the complicated phase diagram, together with an intuitive reason why the easy-axis anisotropy energetically stabilizes the SkX phase at intermediate fields. Depending on the relative strength of the easy-axis anisotropy and the magnetic field, the phase diagram might be divided into three regimes, i.e., [I] the high-field regime where the field exceeds the anisotropy, [II] the low-field regime where the anisotropy exceeds the field, and [III] the medium-filed region where both compete. In the region [I] involving the double-$q$ ($2q,1q$) and the CS ($1q$,U) states, the spin states tend to be ``conical'' induced by the field, while in the region [II] involving the VS ($1q,1q$), the double-$q$ ($2q,2q$), and the single-$q$ ($1q,2q$) states, the spin states tend to be ``vertical'' induced by the anisotropy. Since the conical and vertical states compete with each other, the states in between tend to be virtually ``spherical'', setting the stage for stabilization of the SkX state.

 In summary, by means of extensive MC simulations on the frustrated $J_1$-$J_3$ triangular-lattice Heisenberg model with the easy-axis exchange anisotropy, we have constructed the $T$-$H$ phase diagram containing a rich variety of multiple-$q$ phases. The easy-axis anisotropy stabilizes the triple-$q$ SkX state down to $T=0$ at intermediate fields. As the field gets weaker, the SkX state becomes only metastable, and new multiple-$q$ states with a broken $C_3$ symmetry, the ($2q,2q$) and ($1q,2q$) states, are instead stabilized. In the high-$T$ regime, in addition to the collinear triple-$q$ phase ($Z$ phase), the collinear single-$q$ and double-$q$ states absent in the isotropic model are stabilized by the easy-axis anisotropy.

 Finally, we discuss experimental implications of the present result. Concerning the stability of the SkX and the multiple-$q$ states encompassing it, while the weak easy-axis magnetic anisotropy enhances the SkX formation even at $T=0$, it often accompanies a strong hysteretic effect associated with the $C_3$-breaking. Thus, in order to experimentally clarify the SkX-related phase structure, one needs to examine carefully the possible dependence of the state on the $T$-cooling/$H$-application protocols. Especially when different final states are to be obtained by different protocols to a common ($T,H$), one should determine which state is truly stable. Since the direct comparison of the energies as we did in the present analysis would be difficult experimentally, the long-time off-equilibrium measurements toward equilibrium might eventually be required. Experimental distinction among ($1q,2q$), ($2q,2q$) and ($3q,3q$) from $S({\bm q})$ measurements might sometimes be not easy due to the domain problem, whereas the absence of the topological Hall effect in the former twos could be used as a signature to distinguish them from the SkX state.

 Of course, features of the phase diagram might well depend on the type and the strength of the anisotropy, e.g., the $\gamma$ value, as well as on other perturbative interactions not taken into account in the present model, e.g., the dipolar interaction, quantum fluctuations, higher-order exchange interactions, etc. For example, since the energy difference between the two new $C_3$-broken states, the ($2q,2q$) and ($1q,2q$) states, is rather small, a small change in $\gamma$ and/or other perturbative effects might affect their relative stability.  While further theoretical and experimental studies are desirable to fully clarify the effects of these perturbative interactions, the present work might hopefully serve as a useful starting reference. 

 The author is thankful to Prof. T. Sato, Prof. T. Kurumaji and Dr. K. Mitsumoto for useful discussion. This study was supported by JSPS KAKENHI Grants No.17H06137 and No.24K00572. We are thankful to ISSP, the University of Tokyo, for providing us with CPU time.

\vskip 1cm

\appendix

\section{Definitions of physical quantities}
\label{sec:physical quantities}

In this section of appendix, 
we give definitions of several physical quantities computed in our Monte Carlo (MC) simulations. 

\subsection{Specific heat}
 The specific heat is computed generally via the energy fluctuation. In the vicinity of the first-order transition, to capture the latent-heat contribution, we also compute it via the temperature ($T$) difference of the energy per spin $\langle e\rangle$, i.e., $\Delta \langle e\rangle /\Delta T$, where $\Delta T$ is taken to be 0.001.

\subsection{Spin structure factors}
 The transverse and longitudinal spin structure factors,  $S_{\perp}(\bm{q})$ and $S_{\parallel}(\bm{q})$,  are defined by
\begin{eqnarray}
S_\perp (\bm{q}) &=& \frac{1}{N} \left\langle \sum_{\mu = x,y} \left| \sum_{i=1}^N S_{i\mu} e^{-i \bm{q}\cdot \bm{r}_i}\right|^2  \right\rangle, \label{insta_perp} \\
S_\parallel (\bm{q}) &=& \frac{1}{N} \left\langle \left| \sum_{i=1}^N S_{iz} e^{-i \bm{q}\cdot \bm{r}_i} \right|^2 \right\rangle, \label{insta_para}
\end{eqnarray}
where $N$ is the number of spins, the summation over $i$ is taken over all sites on the triangular lattice, while $\langle \cdots \rangle$ represents the thermal average.

\subsection{Total scalar chirality}
 The local scalar chirality is defined for the upward (downward) triangle by $\chi_{\bigtriangleup(\bigtriangledown)} = \bm{S}_i \cdot \bm{S}_j \times \bm{S}_k~(i,j,k \in \bigtriangleup(\bigtriangledown))$. The total scalar chirality is defined by
\begin{align}
\chi_{\rm tot} &= \frac{1}{2N} \left( \left\langle \left(\sum_{\bigtriangleup} \chi_{\bigtriangleup} + \sum_{\bigtriangledown} \chi_{\bigtriangledown}\right)^2 \right\rangle \right)^{1/2} ,
\end{align}
where the summation $\sum_\bigtriangleup$ ($\sum_\bigtriangledown$) runs over all upward (downward) triangles on the triangular lattice. 

\subsection{Total vector chirality}
 We define the local vector chirality for each upward triangle by ${\bm \kappa}= (\kappa_x,\kappa_y,\kappa_z) = \frac{2}{3\sqrt{3}} \sum_{\langle i,j \rangle} ({\bm S_i}\times {\bm S_j})$, where the summation is taken in the clockwise direction over three bonds on an upward triangle. The transverse and longitudinal components of the total vector chirality per plaquette, $\kappa_t$ and $\kappa_l$, are then defined by
\begin{align}
\kappa_t&= \frac{1}{N} \left( \left\langle \left(\sum \kappa_x \right)^2 +  \left( \sum \kappa_y \right)^2 \right\rangle \right)^{1/2} , \\
\kappa_l&= \frac{1}{N} \left( \left\langle \left(\sum \kappa_z \right)^2 \right\rangle \right)^{1/2}, 
\end{align}
where the summation is taken over all $N$ upward triangles on the triangular lattice.

\begin{figure}[t]
	\centering
	\begin{tabular}{c}
		\begin{minipage}{\hsize}
			\includegraphics[width=\hsize]{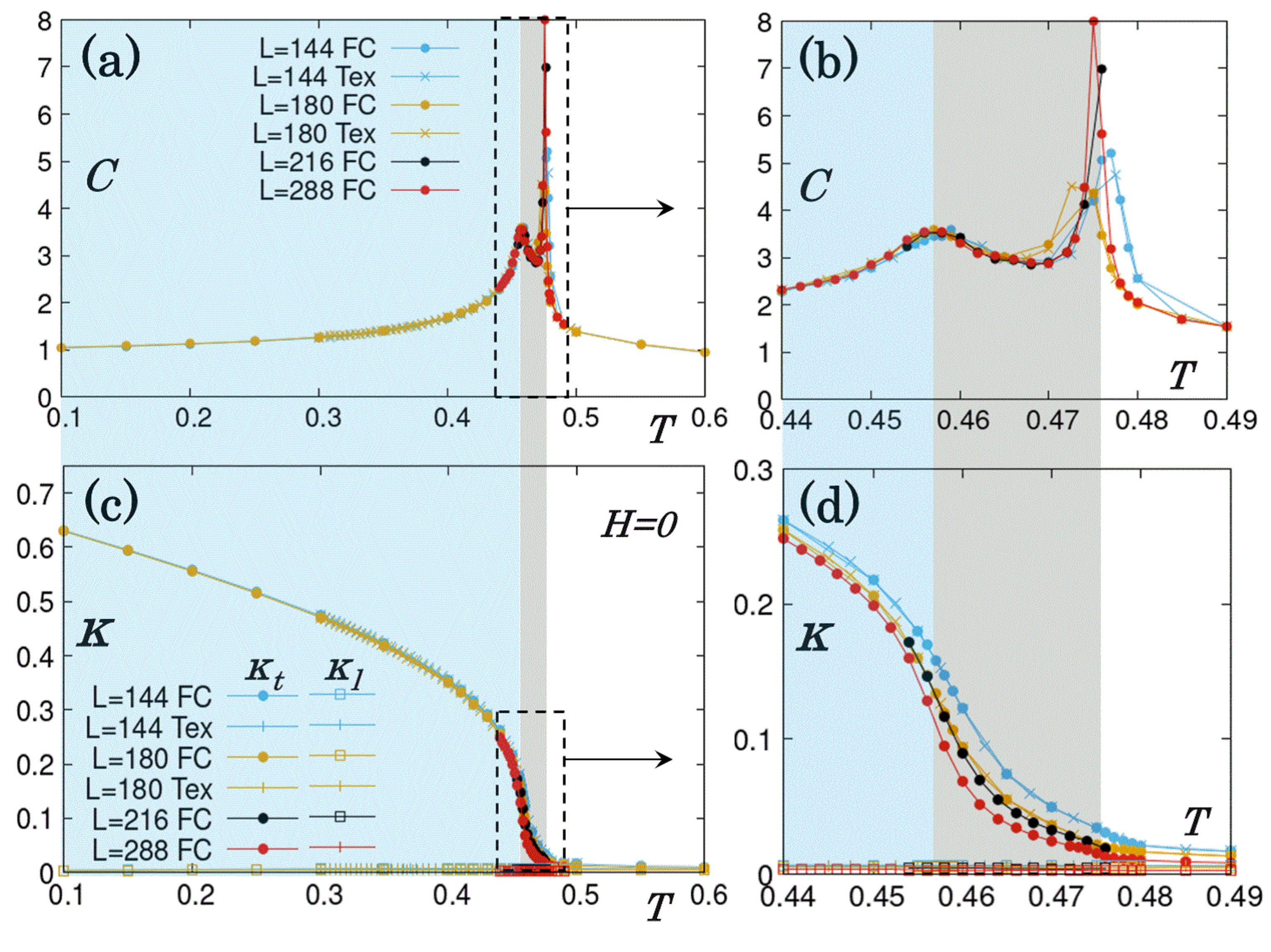}	
		\end{minipage}
	\end{tabular}
	\caption{
The temperature and size dependence of (a) the specific heat, and of (c) the total scalar chirality, in zero field $H=0$. The data of the FC runs and of the temperature-exchange runs ($T_{{\rm ex}}$) are shown. (b) and (d) are magnified views of the transition region depicted by the dashed-line boxes in (a) and (c), respectively. 
	}
	\label{figS1}
\end{figure}

\section{The temperature dependence of physical quantities in magnetic fields}
\label{sec:Tdependence}

 In this section, we wish to show our MC data of the temperature ($T$) dependence of several physical quantities in magnetic fields, which are not shown in the main text.

\subsection{$H=0$}
 
 We begin with the $H=0$ case. At $H=0$, the system exhibits, on decreasing $T$, phase transitions from the paramagnetic to the collinear single-$q$ (D,$1q$)' phase, and then, to the vertical spiral (VS) ($1q,1q$) phase. The quantity which can be regarded as the order parameter of the VS order might be the transverse component of the vector chirality, $\kappa_t$, defined by Eq.(4).

 We show in Figs. 5(a, c) the $T$-dependence of the specific heat and the transverse and longitudinal vector chiralities, respectively. Double peaks of the specific heat  associated with the expected two transitions are observed. The low-$T$ ordered phase is characterized by a nonzero $\kappa_t$ with a vanishing $\kappa_l$ defined by Eq.(5), consistently with the VS ordering. In the intermediate (D,$1q$)' phase, $\kappa_t$ tends to decrease systematically with the system size $L$, consistently with the expected collinear ordering. Indeed, the spin structure factor of the intermediate phase shown in Figs.10(e, f)  below also supports the (D,$1q$)' nature of the intermediate phase.

\begin{figure}[t]
	\centering
	\begin{tabular}{c}
		\begin{minipage}{\hsize}
			\includegraphics[width=\hsize]{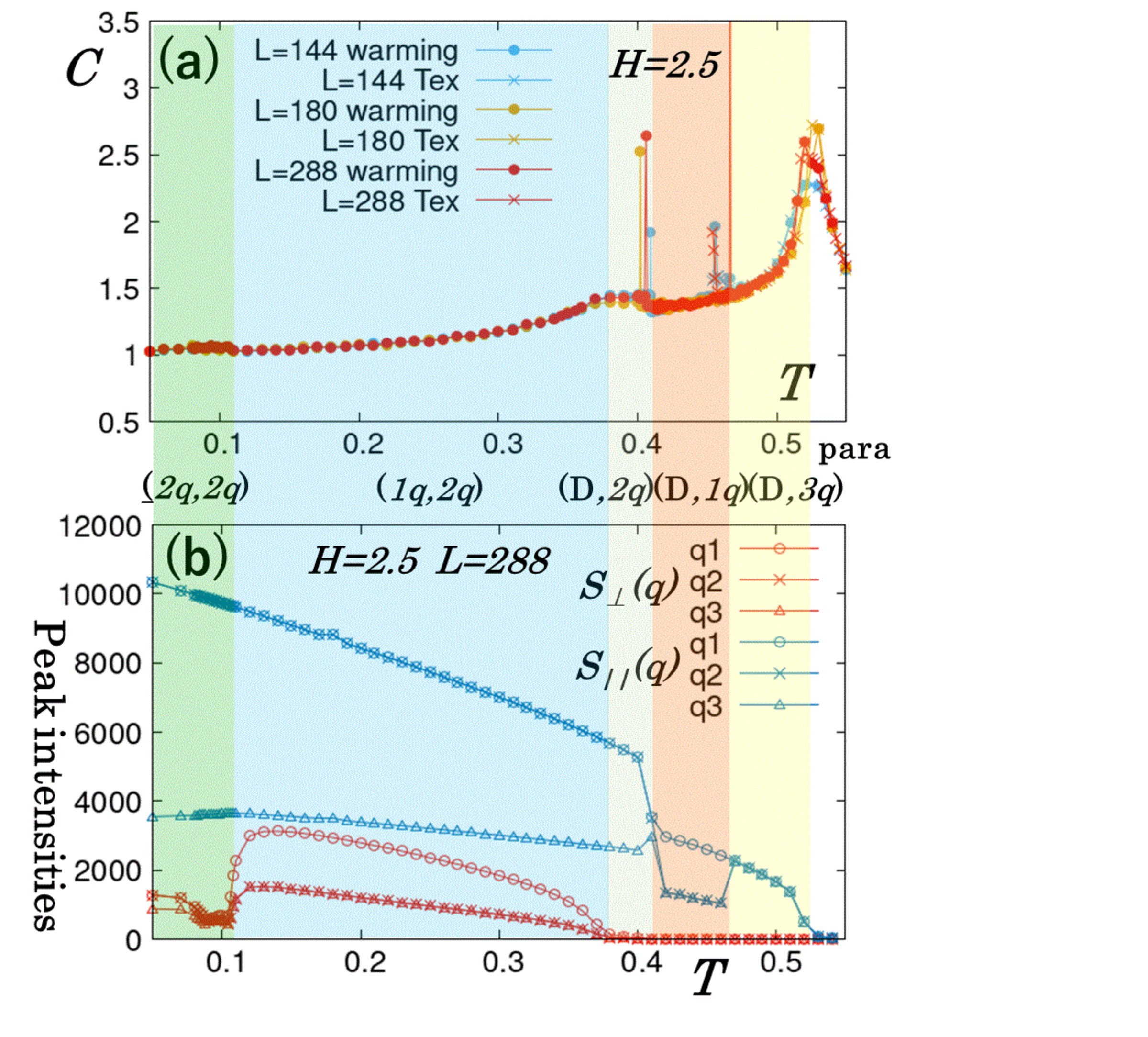}	

		\end{minipage}
	\end{tabular}
	\caption{
The temperature dependence of (a) the specific heat for various sizes, and of (b) the intensities of the three (quasi-)Bragg peaks of $S_\perp({\bm q})$ and $S_\parallel({\bm q})$ for $L=288$ which are ordered according to their intensities. The field is $H=2.5$. The data are taken by the gradual warming runs from the $T=T_0$ state prepared by the ZFC run as explained in the main text. In (a), the data taken by the temperature-exchange runs ($T_{{\rm ex}}$) are also shown.
	}
	\label{figS2}
\end{figure}

\subsection{$H=2.5$}

  To demonstrate the richness of the phase structure of the model, we have shown in Figs. 4(a, b) of the main text the $T$-dependence of the specific heat and the intensities of the (quasi-)Bragg peaks of the spin structure factor at a representative field $H=2$, which exhibits a variety of multiple-$q$ phases, i.e., the ($2q,2q$), ($1q,2q$), (D,$2q$), (D,$1q$), (D,$3q$) states on increasing $T$ before finally reaching the paramagnetic state.  In this subsection, we show similar plots for a different field $H=2.5$. In Figs. 6(a, b), we show respectively the $T$-dependences of the specific heat and of the intensities of the three relevant (quasi-)Bragg peaks ${\bm q}_i^*$ ($i=1,2,3$) of the transverse and longitudinal spin structure factors $S_\perp({\bf q})$ and $S_\parallel({\bf q})$, which are ordered  according to their intensities. The data are taken by the gradual warming runs from the $T=T_0$ state prepared by the ZFC run as explained in the main text. Similar phase sequences as observed in Figs. 4(a,b) of the main text for $H=2$ are also observed for $H=2.5$.

\begin{figure}[t]
	\centering
	\begin{tabular}{c}
		\begin{minipage}{\hsize}
			\includegraphics[width=\hsize]{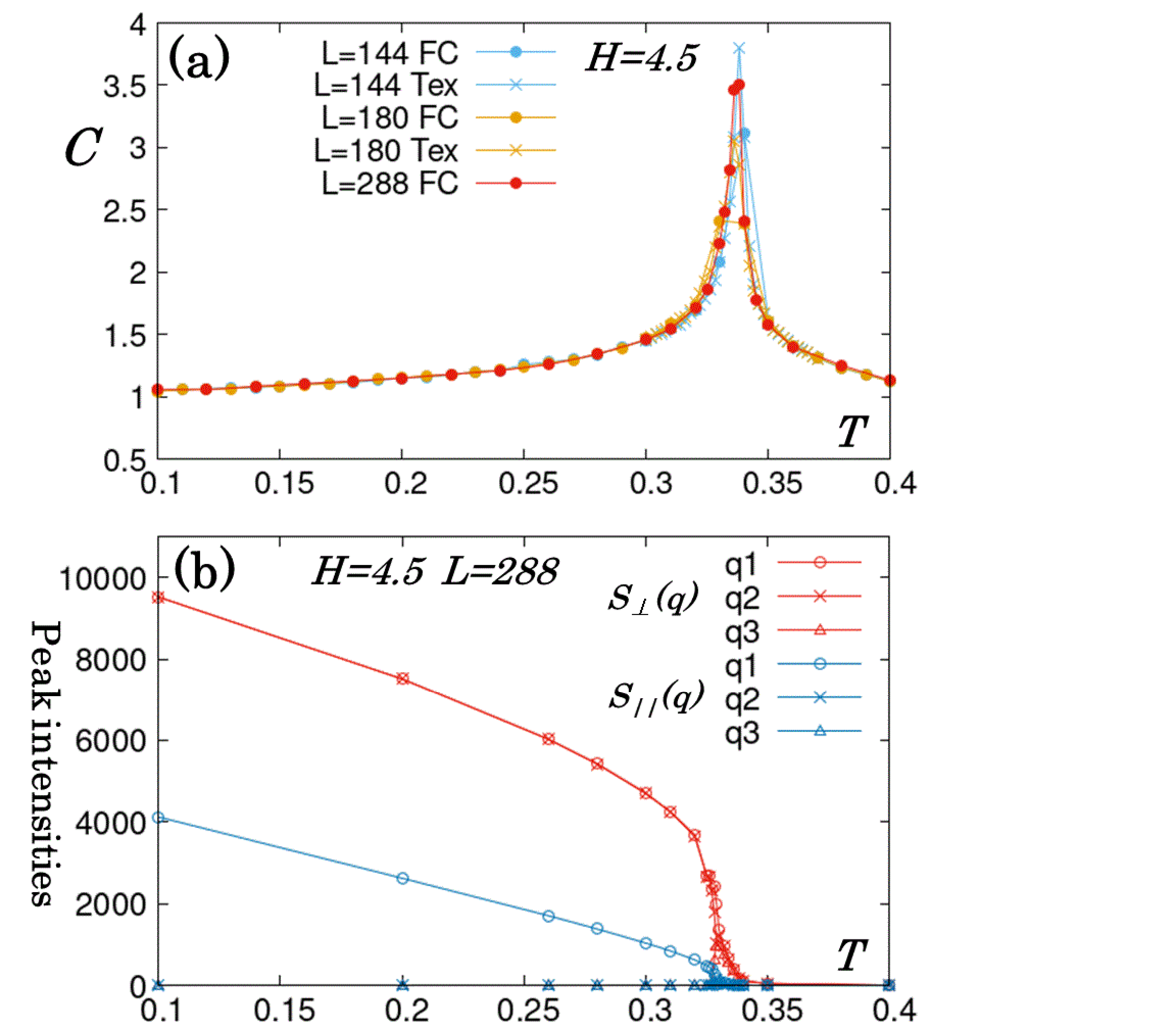}	
		\end{minipage}
	\end{tabular}
	\caption{
The temperature dependence of physical properties at $H=4.5$. (a) The specific heat for various sizes, and of (b) the intensities of the three (quasi-)Bragg peaks of $S_\perp({\bm q})$ and $S_\parallel({\bm q})$ for $L=288$ which are ordered according to their intensities. The data are taken by the FC runs, while in (a) the data taken by the temperature-exchange runs ($T_{{\rm ex}}$) are also shown. 
	}
	\label{figS3}
\end{figure}

\subsection{$H=4.5$}

 At a higher field $H=4.5$, the system exhibits on decreasing $T$ a phase transition from the paramagnetic state to the double-$q$ ($2q,1q$) state. The spin and the vector-chirality configurations at $H=4.5$ have been given in Figs. 4(e, f) of the main text. We show in Figs. 7(a, b) the $T$-dependence of the specific heat and the intensities of the three (quasi-)Bragg peaks of the transverse and longitudinal spin structure factors $S_\perp({\bm q})$ and $S_\parallel({\bm q})$, at $H=4.5$. The occurrence of a single transition to the ($2q,1q$) ordered state can be seen from the figure.

\begin{figure}[t]
	\centering
	\begin{tabular}{c}
		\begin{minipage}{\hsize}
			\includegraphics[width=\hsize]{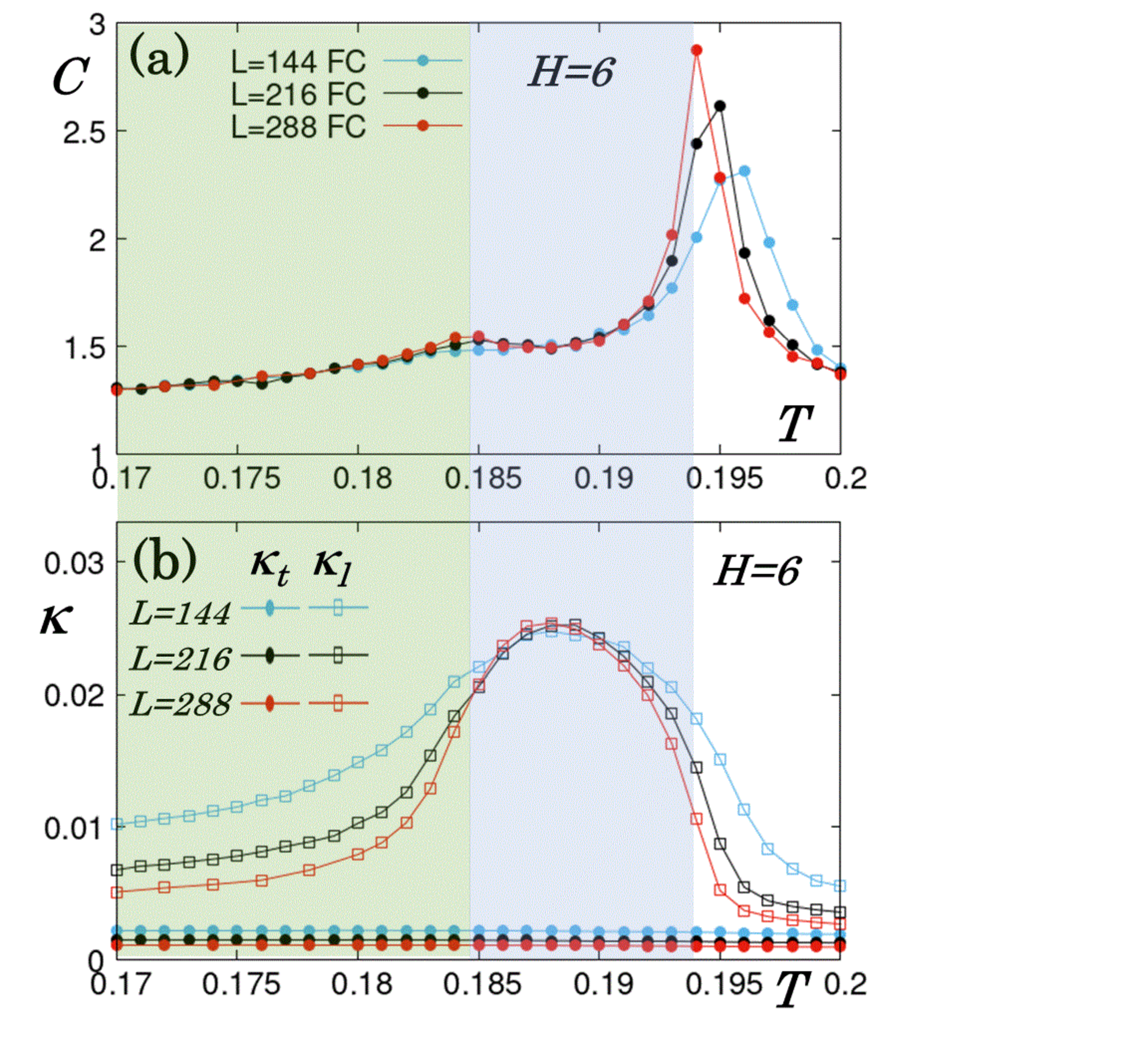}	

		\end{minipage}
	\end{tabular}
	\caption{
The temperature dependence of physical properties in the transition region at $H=6$: (a) The specific heat, and (b) the transverse and longitudinal vector chiralities, $\kappa_t$ and $\kappa_l$, for various sizes. The data are taken by the FC runs. 
	}
	\label{figS4}
\end{figure}

\subsection{$H=6$}

 At a still higher field $H=6$, the conical spiral (CS) ($1q$,U) state intervenes the paramagnetic and the ($2q,1q$) states, as can be seen from the $H$-$T$ phase diagram shown in Fig. 1 of the main text. To demonstrate this, we show in Figs. 8(a, b) the $T$-dependence of the specific heat and of the transverse and longitudinal vector chiralities, $\kappa_t$ and $\kappa_l$. Note that the CS state is characterized by a nonzero $\kappa_l$ with a vanishing $\kappa_t$, in contrast to the VS state characterized by a nonzero $\kappa_t$ with a vanishing $\kappa_l$. The specific heat shown in Fig. 8(a) now exhibits double peaks, suggesting the appearance of an intermediate phase. As can be seen from Fig. 8(b), the intermediate state is characterized by a nonzero $\kappa_l$ with a vanishing $\kappa_t$, consistently with the CS nature of the state. We note that the CS state intervening the paramamagnetic and the ($2q,1q$) states have also been observed in the isotropic model \cite{OkuboChungKawamura}.

\begin{figure}[t]
	\centering
	\begin{tabular}{c}
		\begin{minipage}{\hsize}
			\includegraphics[width=\hsize]{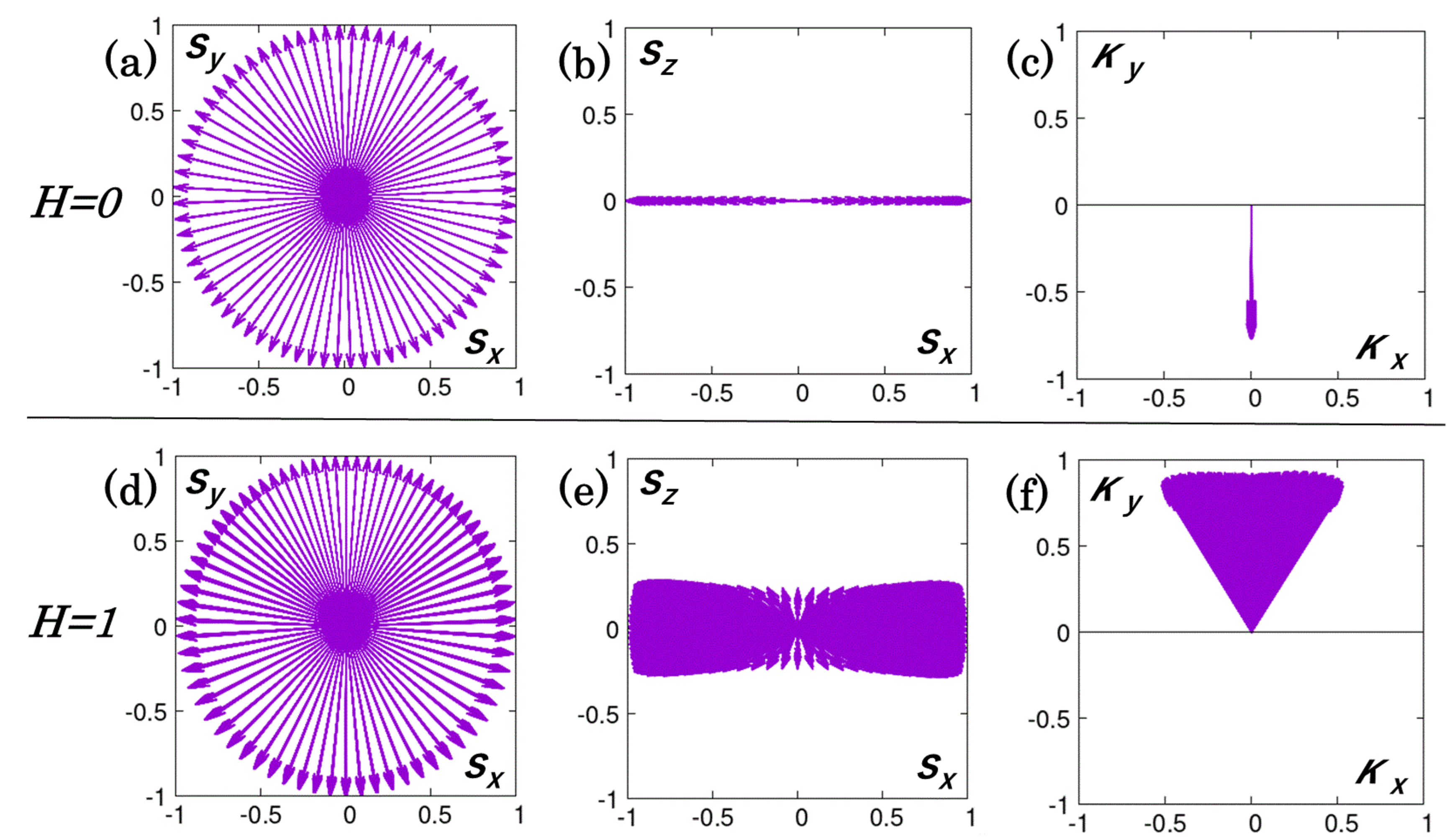}	
		\end{minipage}
	\end{tabular}
	\caption{
The projected spin (a,d) ($S_{x},S_{y}$) and (b,e) ($S_{x},S_{z}$) components, and (c,f) the vector-chirality ($\kappa_x,\kappa_y$) components in the VS ($1q,1q$) phase state at $T=0.0001$. The upper row (a,b,c) corresponds to $H=0$, and the lower row (d,e,f) corresponds to $H=1$. The lattice size is $L=180$. 
	}
	\label{figS5}
\end{figure}

\begin{figure}[t]
	\centering
	\begin{tabular}{c}
		\begin{minipage}{\hsize}
		\includegraphics[width=\hsize]{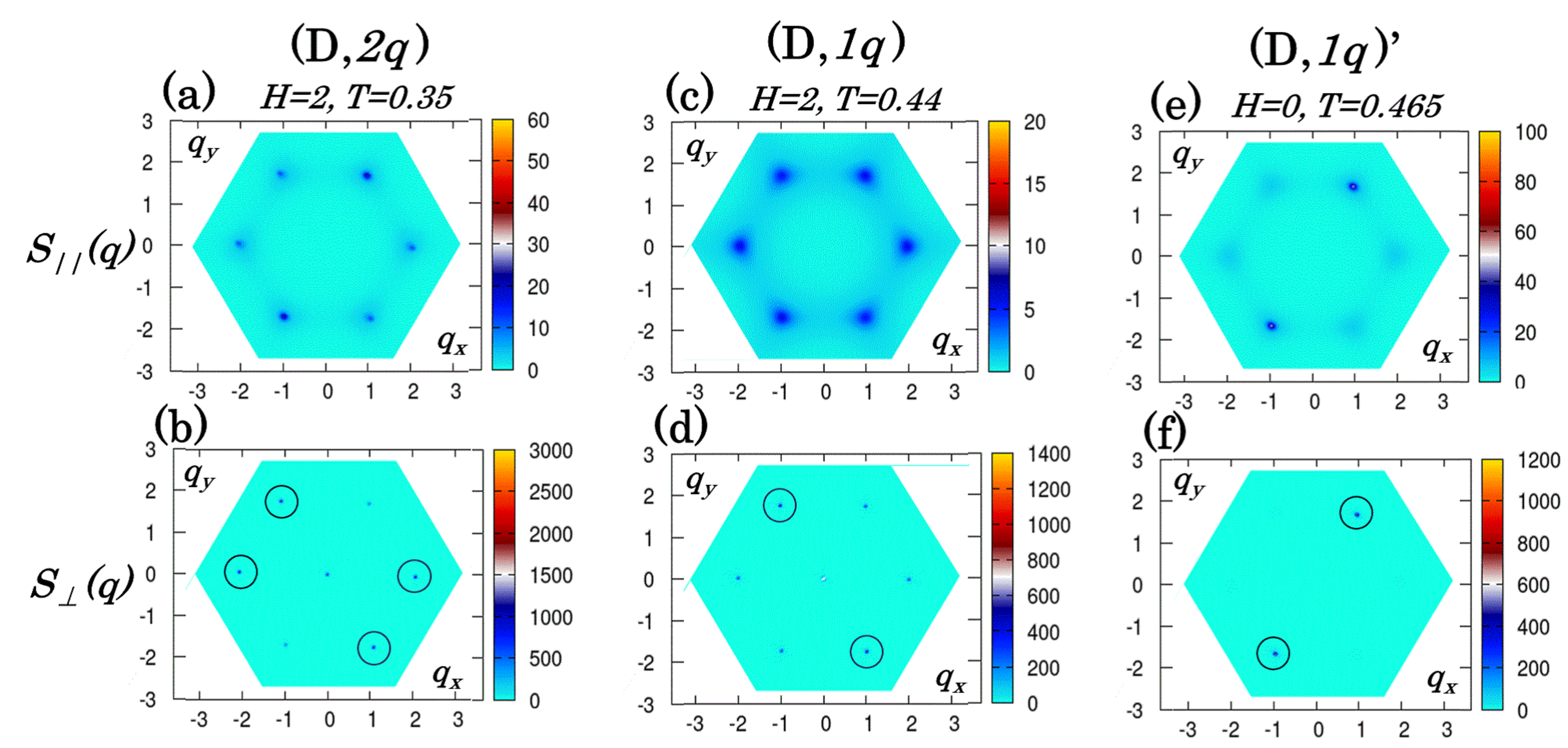}	
		\end{minipage}
	\end{tabular}
	\caption{
 The transverse and longitudinal spin structure factors, $S_\perp({\bm q})$ and $S_\parallel({\bm q})$, in [left] the (D,$2q$) phase at $H=2$ and $T=0.35$, [middle] the (D,$1q$) phase at $H=2$ and $T=0.44$, and [right] the (D,$1q$)' phase at $H=0$ and $T=0.465$. The upper row (a,c,e) represents $S_\perp({\bm q})$, while the lower row (b,d,f) represents $S_\parallel({\bm q})$. The lattice size is $L=180$.
	}
	\label{figS6}
\end{figure}

\section{ Properties of each ordered phase}
\label{sec:eachphase}

 In this section, we show some of the properties of each ordered phase not given in the main text.

\subsection{The ($1q,1q$) vertical-spiral phases}

 We begin with the VS state in zero and lower fields. In Fig. 9, we show the projected plots of the spin (a) ($S_x,S_y$) and (b) ($S_x,S_z$), and (c) the vector chirality ($\kappa_x,\kappa_y$) configurations. As can be seen from Figs. 9(a-c), the VS state at $H=0$ is a vertical coplanar state with a definite rotation. At a finite field $H=1$, by contrast, the VS becomes noncoplanar as can be seen from Figs. 9(d-f), as briefly mentioned in the main text. Yet, the state does not exhibit a counter rotation exhibited by the ($1q,2q$) state as explained in the main text. Compare Figs. 9(c, f) with Fig. 4(c) of the main text.

\subsection{The collinear (D,$1q$), (D,$2q$) phases}

 As mentioned in the main text, the present anisotropic model also exhibits at higher temperatures the collinear phases not realized in the corresponding isotropic model, including the two types of collinear single-$q$ phase, which are called (D,$1q$) and (D,$1q$)' phases each stabilized in intermediate- and low(zero)-field regimes, and the collinear double-$q$ (D,$2q$) phase. In Fig. 10, we show the transverse and longitudinal spin structure factors, $S_\perp({\bm q})$ and $S_\parallel({\bm q})$, for these three collinear phases, i.e., (D,$2q$), (D,$1q$) and  (D,$1q$)' phases.

\begin{figure}[t]
	\centering
	\begin{tabular}{c}
		\begin{minipage}{\hsize}
			\includegraphics[width=\hsize]{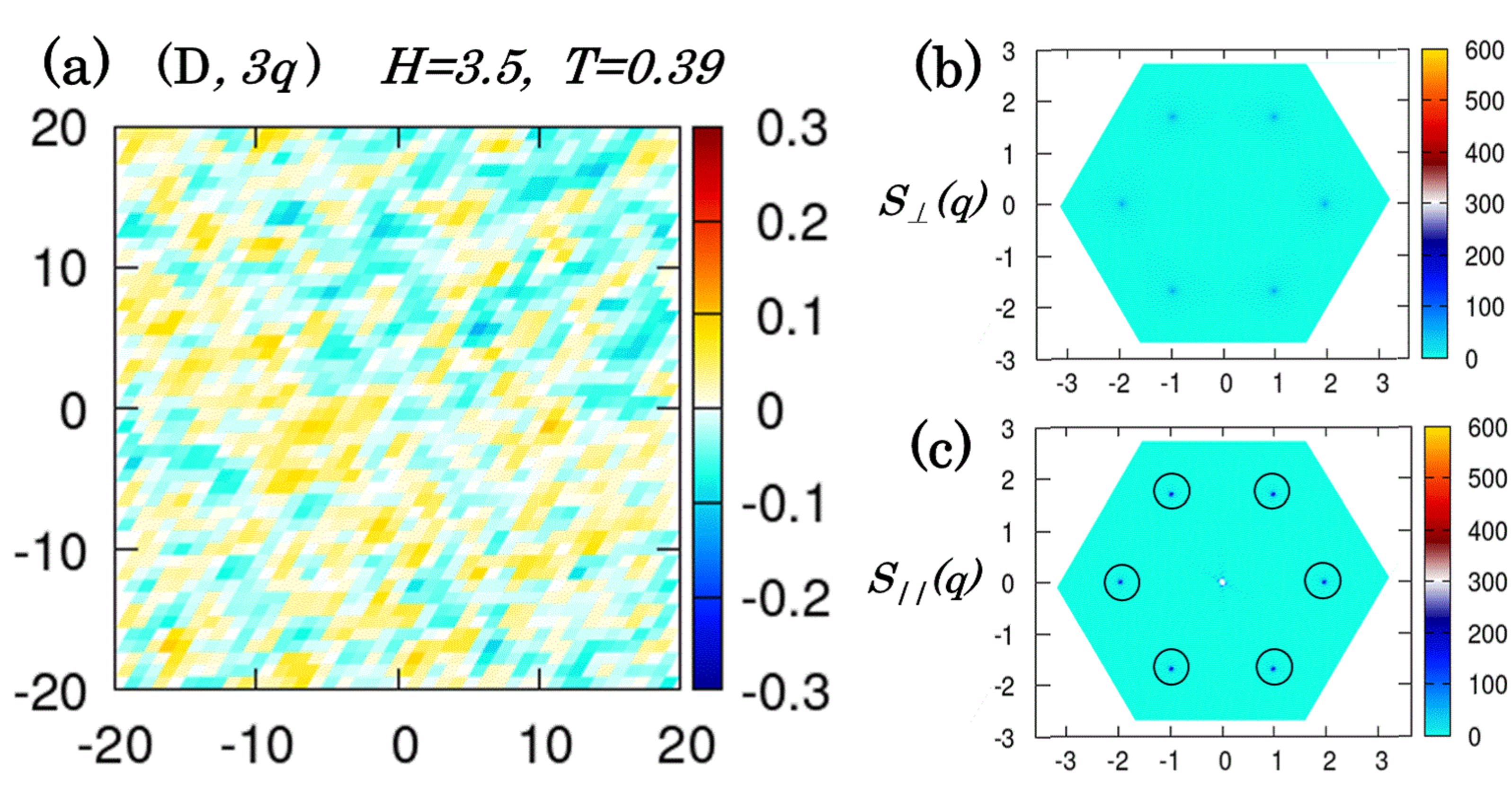}	
		\end{minipage}
	\end{tabular}
	\caption{
The real-space scalar-chirality configuration and the spin structure factors in the $Z$ (D,$3q$) state at $H=3.5$ and $T=0.39$. (a) The color map of the typical real-space configurations of  the scalar chirality, and the transverse and longitudinal spin structure factors, (b) $S_\perp({\bm q})$, and (c) $S_\parallel({\bm q})$. The lattice size is $L=180$, while a part of the lattice is shown in (a).
	}
	\label{figS7}
\end{figure}

\subsection{The collinear (D,$3q$) phase}

 The $Z$ phase, i.e., the collinear triple-$q$ (D,$3q$) state, which also exists in the isotropic model, appears in the anisotropic model, too. In fact, the anisotropy enhances the stability of this phase considerably in the $T$-$H$ phase diagram \cite{OkuboChungKawamura}, while the fundamental character of the phase remains the same as in the isotropic case. To demonstrate the random-domain character consisting of the SkX and anti-SkX states, we show in Fig. 11(a) a typical real-space configuration of the scalar chirality. In Figs. 11(b,c), we show the corresponding spin structure factors, $S_\perp({\bm q})$ and $S_\parallel({\bm q})$, respectively.


\begin{thebibliography}{99}
\bibitem{mqnq} For example, if $S({\bm q})$  possesses three (quasi-)Bragg peaks at ${\bm q}_1$, ${\bm q}_2$, and ${\bm q}_3$ with the relation $S({\bm q}_1)=S({\bm q}_2)>S({\bm q}_3)>0$, we call it $2q$.
\bibitem{qandS} Note that the real-space ($x,y$) (or the ${\bm q}$-space ($q_x,q_y$)) part and the spin-space ($S_x,S_y$) part are totally uncoupled in the present model so that any spin configuration generated by aribitrary global spin rotation around  the magnetic-field ($S_z$) axis is equally possible.  In Figs. 3(c) and (d), just for illustation, we take a special choice that the coplanar spin place coincides with the ($S_x,S_z$) plane. The same convention is employed also in Fig. 4(c).
\bibitem{comment} The ($2q,2q$) $\leftrightarrow$ ($1q,2q$) transition temperature is estimated to be $T_c\simeq 0.23$ from Fig.4(a) and (b). Even within the ($2q,2q$) phase, however, some step-like ``structure'' or ``weak anomaly'' exists in Fig.4(b) just below $T_c$, while the basic ($2q,2q$) character of $S({\bm q})$ remains the same. We regard such ``structure'' only as secondary one associated with, e.g., the possible incommensurability effect or the domain formation, although we cannot completely rule out the possibility that the another phase with the same ($2q,2q$) ordering pattern intervenes between the ($2q,2q$) and  ($1q,2q$) phases just below $T_c$.
\end{thebibliography}
\end{document}